\documentclass[twocolumn]{aa}
\usepackage{graphicx}
\def\HI {H\kern0.1em{\sc i}} 
\def\HII {H\kern0.1em{\sc ii}}

\def\kms{km s$^{-1}$}
\def\solmass {\hbox{M$_{\odot}$}}
\def\solum {\hbox{L$_{\odot}$}}

\begin{document}
   \title{The \HI\ content of the recently discovered field dwarf galaxy \object{APPLES~1}}


   \author{A.\ Tarchi\inst{1,2}
          \and
          J.\ Ott\inst{3}
          \and
          A.\ Pasquali\inst{4}
          \and
          I.\ Ferreras\inst{5}
          \and
          P.\ Castangia\inst{2,6}
          \and
          S.\ S.\ Larsen\inst{7}
          }

   \offprints{A. Tarchi, \email{a.tarchi@ira.cnr.it}}

   \institute{INAF - Istituto di Radioastronomia, 
              via Gobetti 101, I--40129 Bologna, Italy
          \and
              INAF - Osservatorio Astronomico di Cagliari,
              Poggio dei Pini, Strada 54, I--09012 Capoterra (CA), Italy 
          \and
              CSIRO Australia Telescope National Facility, Cnr Vimiera \& Pembroke Roads, Marsfield NSW 2122, Australia 
          \and
              Institute of Astronomy, ETH, H\"onggerberg, HPF, 8093 Z\"urich, Switzerland
           \and
Department of Physics and Astronomy, University College London, Gower St., London WC1E 6BT, United Kingdom
           \and
              Universit\`a di Cagliari, Dipartimento di Fisica, Cittadella Universitaria, 09042 Monserrato (CA), Italy
          \and 
           European Southern Observatory, Karl-Schwarzschild-Strasse 2, D-85748 Garching, Germany}

   \date{Received ; accepted}

   \abstract{We present observations in a search for neutral hydrogen associated with the recently detected field dwarf galaxy \object{APPLES~1}, performed with the Parkes radiotelescope. The observed radio spectrum shows no evident ($>$ 3$\sigma$ rms) line emission indicating an upper limit for the \HI\ content of the galaxy of $\sim 10^{6}\:\solmass$ and providing an upper value for the M$_{\rm HI}$/L$_{\rm B}$ ratio equal to 2.4 \solmass /\solum. The low value of the \HI\ content suggested by the observations, together with the galaxy optical morphology, might indicate that \object{APPLES~1} is a dwarf spheroidal. This indication is in contrast with the evidence of recent star formation, which is typical for dwarf irregular galaxies. This may suggest that \object{APPLES~1} belongs to the class of mixed dwarf irregular/spheroidal transition-type galaxies. We also conclude that the relatively low neutral gas mass in \object{APPLES~1} can be explained by an extended and inefficient star formation process, without the need for a dramatic event such as enhanced star formation or a past encounter with a massive galaxy or galaxy group.

   \keywords{Galaxies: individual: \object{APPLES~1} -- Galaxies: dwarf -- 
          Radio lines: ISM -- Radio lines: galaxies
               }
   }

\titlerunning{The \HI\ content of the dwarf galaxy \object{APPLES~1}}
\authorrunning{A.\ Tarchi et al.}

   \maketitle
%

\section{Introduction}

Dwarf galaxies, i.e. galaxies with low mass, low luminosity, low metallicity, 
and small in size, constitute the most numerous class of galaxies in the Universe. In the hierarchical galaxy formation framework, dwarf galaxies (or, in general, small dark matter halos) are the building blocks of larger structures and, therefore, should have been even more numerous in the early Universe. Accordingly, the local population of dwarf galaxies may be considered as survivors of an originally much richer population. In the Local Group, far fewer dwarf galaxies at the low-mass end have been observed than the number of predicted surviving dark matter halos (e.g. Klypin et al.\ \cite{klypin}). This may imply that we are either missing a large fraction of low-mass galaxies, that these halos do not have detectable counterparts, or that the aforementioned theory of galaxy formation needs to be improved. 

Detailed studies of nearby dwarf galaxy populations are therefore vital to understand the formation and evolution of galaxies in general. The best-studied dwarf galaxies are in the Local Group but in recent years more and more results are reported also on dwarf galaxies belonging to other groups or clusters of galaxies, e.\ g.\ the Sculptor, M~81, and Cen A groups. Dwarf galaxies are classified in several classes (for a review on dwarf galaxies, see e.\ g.\ Mateo \cite{mateo}). These classes are distinguished by different morphological characteristics and physical quantities. Among the latter, the M$_{\rm HI}$/L$_{\rm B}$ ratio (hereafter given in units of \solmass /\solum) is one of the most widely used and provides an indication of the abundance of gas w.r.t. the recent star formation activity in the galaxy. The main three classes of dwarf galaxies are: (a) dwarf irregulars (dIrrs), that are irregular in appearance and seem to exhibit scattered bright \HII\ regions in the optical. dIrrs are gas-rich and the morphology of the neutral gas phase, traced by hydrogen (\HI), is dominated by prominent shells, bubbles, and individual gas clumps. dIrrs have been so far found in clusters, groups, and in the field; (b) Dwarf ellipticals (dEs), that are compact, spherical or elliptical in appearance. They tend to be found close to massive galaxies, usually have little or no detectable gas; (c) dwarf spheroidals (dSphs), that are similar to dEs but with a more diffuse morphology, with low-surface brightness, and very little central concentration. They are the faintest, least massive galaxies known and are usually found in close proximity to massive galaxies and, similarly to dEs, are gas-poor objects. It has been proposed that dSphs were initially gas rich but that the gas was removed by tidal interactions within their group or cluster environment (e.\ g.\  van den Bergh \cite{vandenbergh}). While dIrrs typically have M$_{\rm HI}$/L$_{\rm B}$ ratios greater than unity, the values for the other two classes (dEs \& dSphs) range from 1 down to 0.001.

Several studies propose an evolutionary scenario between these classes of objects according to which gas removal processes would turn a gas-rich dIrr into a gas-poor dSph/dEs (e.\ g.\ van Zee \cite{vanzee}; Grebel, Gallagher \& Harbeck \cite{grebel}, hereafter GGH). Among these processes, external mechanisms like tidal and ram-pressure stripping are usually preferred to internal ones (such as supernova explosions, stellar winds or simple gas consumption due to star formation) since gas-poor dwarfs are found closer to the group centers and are therefore more susceptible to external forces (e.\ g.\ Mayer et al.\ \cite{mayer}). However, some gas-deficient dSph-like galaxies (e.\ g.\ \object{LGS~3}, \object{Phoenix}, \object{Cetus} and \object{Tucana}) have been found relatively far from any large galaxy. In particular, in \object{LGS~3} and \object{Phoenix} small but detectable amount of \HI\ was revealed (Carignan, Demers, \& Cot\'{e} \cite{carignan}; Lo, Sargent, \& Young \cite{lo}) providing a total \HI\ mass for each galaxy of few 10$^{5}$ \solmass, while in \object{Tucana}, no significant \HI\ (M$_{\rm HI}$ $<$ 1.5 $\times$ 10$^{4}$ \solmass) was detected within the optical boundary of the galaxy (Oosterloo, Da Costa, \& Staveley-Smith \cite{oosterloo}). Because of relative isolation of these `transition-type dwarfs' (e.\ g.\ Sandage \& Hoffman \cite{sandage}; GGH), their low gas mass cannot be easily due to the aforementioned external mechanisms and it has to depend on internal star formation processes. Even so, the transition from rotating dIrrs to non-rotating dSph/dEs also implies an angular momentum dissipation that is difficult to obtain without external forces, like tidal ones, an issue that still needs to be solved and deserves exploration. The detection of genuine field dSphs galaxies (i.\ e.\ clearly isolated galaxies) and detailed studies of these object would then be even more suitable to investigate possible causes of gas depletion and the 'dissipation' problem. Unfortunately, so far, the detection of such a galaxy has been quite rare. 

Very recently, within the framework of the ACS (Advanced Camera for Surveys on board {\it {HST}}) Pure Parallel Ly$\alpha$ Emission Survey program (APPLES), a candidate dSph galaxy has been discovered in the field (\object{APPLES~1}) at a distance of about 9.6 Mpc (Pasquali et al.\ \cite{pasquali}; hereafter PLF). It is the faintest field dwarf known so far (with M$_V \simeq$ -9) and is very likely not associated with any galaxy group or cluster (PLF). Its luminosity-weighted age\footnote{The age of the galaxy is clearly indicated by the Balmer break and adjacent continuum slope. These parameters vary as a function of the age of the stellar population which emit the most at those wavelengths (3800 $-$ 5000 \AA). Therefore, the age derived from these features is said to be weighted by the luminosity of the stars which contributed most of the light in the blue part of the spectrum.} is of about 3 Gyr and its observed spectral energy distribution (SED) suggests that \object{APPLES~1} is forming stars at a slow pace. Since \object{APPLES~1} is an isolated system, all mechanisms that extend its star formation history must be intrinsic. Therefore, the presence of a spherical-like morphology (similar to dSphs) but also of recent star formation (usually found in dIrr) and its isolation makes \object{APPLES~1} an interesting object where to investigate the relation between the two different classes of dwarf galaxies and an interesting test case for studies of the formation and evolution of unperturbed dwarf galaxies (PLF).

In this article, we investigate the \HI\ content of \object{APPLES~1}. In Sect.~2 we describe the observations and data reduction and our results are presented in Sect.~3. This is followed in Sect.~4 by a concise discussion on the nature of \object{APPLES~1} and on possible mechanisms that could justify its relatively low gas mass.    

\section{Observations and data reduction}

We observed \object{APPLES~1} at the frequency of the \HI\ line on March 7, 2005, for a total on-source time of 7.5 hours using the L-Band multi-beam receiver on the 64\,m ATNF Parkes telescope\footnote{The Parkes telescope is part of the Australia Telescope which is funded by the Commonwealth of Australia for operation as a National Facility managed by CSIRO.}. The full width to half power beamwidth of each horn is 14\farcm4. We used an 8 MHz bandwidth centered on the systemic velocity of the galaxy ($V_{hel}$ = (674 $\pm$ 30) \kms; PLF), split in 2048 channels, thus yielding a velocity resolution of $\sim$ 0.8 \kms. Both linear polarizations were observed.

The data reduction was performed in two steps. The first involved the task LIVEDATA, and consisted of flagging channels at the edges of the band and those where Galactic \HI\ emission is present (in our broad-band spectra \HI\ emission was found at a velocity of $\sim$ 0 \kms, and hence, confidently not produced by \object{APPLES~1}) smoothing the remaining channels, and subtracting the baseline (a polynomial of 3rd order was used). The second step made use of the imaging program GRIDZILLA that allowed us to produce the final, combined spectrum after having weighted the individual spectra by the inverse square of the system temperature (T$_{sys}$) and accounted for radio frequency interferences (RFI) of artificial origin.

\section{Results}

Fig.~1 shows the \HI\ spectrum of \object{APPLES~1} taken with the Parkes radiotelescope and smoothed down to a resolution of 8.3 \kms\ (to obtain a better signal-to-noise ratio). The velocity range is $\sim$ 1300 \kms. No significant \HI\ feature is detected within the observed velocity range down to the 3$\sigma$ rms noise level per channel of 5.22 mJy.  Under the assumption that \object{APPLES~1} is not rotating and that we are calculating the mass within a 8 \kms\ channel (the typical velocity dispersion of a dwarf galaxy ranges between 5 and 13 \kms; Stil \& Israel \cite{stil}), our observations result in a 3$\sigma$ upper limit for the \HI\ mass of $\sim$ 10$^{6}$ \solmass. However, there is a hint for a possible line feature at a velocity of $\sim$ 695 \kms. So far, the only available measurement of the recessional velocity of the galaxy is the one reported by PLF, as measured from its absorption Balmer lines, and amounting to $V_{hel}$ = (674 $\pm$ 30) \kms. The velocity of this weak \HI\ feature is thus within the errors, and makes it possible to be consistent with neutral hydrogen gas emission from \object{APPLES~1}. Assuming that this feature is real, the corresponding \HI\ mass for a 4.5 mJy peak and a 8 \kms\ channel width would be 7.8 $\times$ 10$^{5}$ \solmass. However, due to the weakness of the signal, we prefer not to take this tentative feature into account in the following discussion.

\begin{figure}
\includegraphics[width= 8.8 cm]{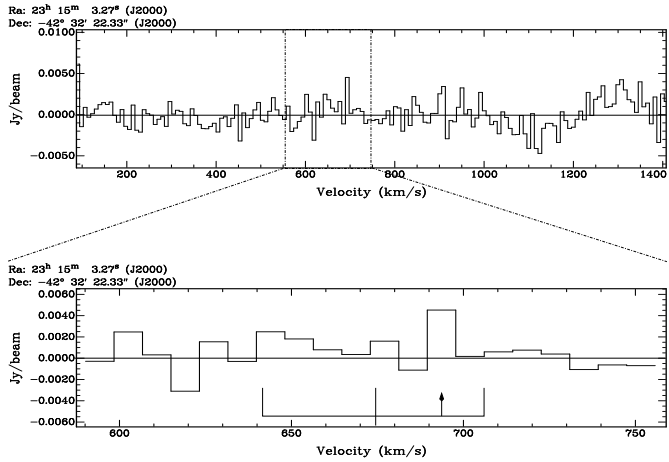}
\caption{Full velocity range (upper panel) and close-up (lower panel) \HI\ Parkes spectrum of \object{APPLES~1}, which has been smoothed by a factor of ten. The velocity resolution is 8.3 \kms\ and the rms (1$\sigma$) per channel is 1.74 mJy/beam. The beam (HPBW) is $\sim$ 14$^{\prime}$. The thick lines in the lower panel indicates the recessional velocity of the galaxy together with its uncertainty, $V_{hel}$ = (674 $\pm$ 30) \kms. The arrow marks the weak feature possibly due to \HI\ emission from \object{APPLES~1}.}
\label{fig1}
\end{figure}

\section{Discussion}

As already anticipated in Sect.\ 1, plotting the \HI\ masses versus distance from closest massive galaxies or galaxies group centers, it has been shown that while most dSphs and dEs are located in the vicinity of major galaxies, dIrrs are found at larger distances (e.\ g.\ van den Bergh \cite{vandenbergh}; GGH). Following this scenario, isolated galaxies like \object{APPLES~1} should have properties typical of dIrrs. Based on the ACS images, PLF tentatively classified \object{APPLES~1} as a dSph. The absence of nebular emission suggests no significant current star formation activity. The relatively low \HI\ content derived from our observation would in principle support a dSph type. On the other hand, the presence of strong absorption Balmer lines in \object{APPLES~1} and of O and B stars suggest that star formation was taking place until quite recently (PLF). This fact seems to rule out the dSph nature and to favour a dIrr classification. Hence, in order to understand the true nature of our target, we find useful to relate it with the other known dwarf galaxies.

\begin{figure}
\includegraphics[width= 8.8 cm]{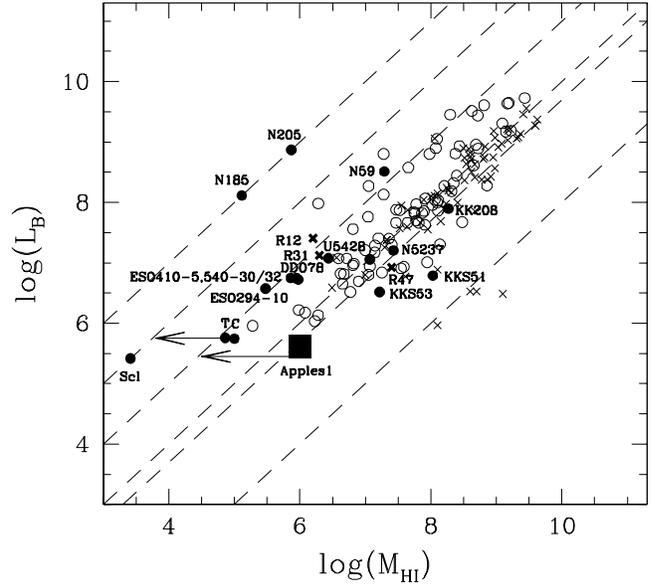}
\caption{Filled black square: \object{APPLES~1}. Open circles: Local Group (Mateo \cite{mateo}), Cen A, Sculptor, M81, M83, IC342, Maffei, Canes Venatici I groups (Huchtmeier et al.\ \cite{huchtmeier}; Karachentsev et al.\ \cite{karachentsev2}; Paturel et al.\ \cite{paturel}; Karachentsev \cite{karachentsev1}; Bouchard et al.\ \cite{bouchard}). Crosses: field dwarf galaxies (van Zee \cite{vanzee}; Roberts et al.\ \cite{roberts}). All dSphs/transition-type dwarfs are represented with filled or bold symbols and labelled with their names (to avoid confusion the names of the dSphs field galaxies taken from Roberts et al.\ \cite{roberts} are shortened using an ``R'' followed by the number of the source reported in their Table~1). The dashed lines correspond to a constant M$_{\rm HI}$/L$_B$ ratio of 100, 2, 1, 0.1, 0.01, and 0.001 from right to left.}
\label{fig2}
\end{figure}

Using a value for the luminosity of \object{APPLES~1} in B band of 4.2 $\times$ 10$^{5}$ \solum, as derived from the best-fitting SED of the galaxy (PLF), we derive a ratio M$_{\rm HI}$/L$_{\rm B}$ smaller than 2.4. In Fig.~\ref{fig2} \HI\ masses are plotted as a function of B-band luminosities for different types of dwarf galaxies and environments. Superimposed are the constant M$_{\rm HI}$/L$_{\rm B}$ (dashed) lines for ratio values (from right to left) of 100, 2, 1, 0.1, 0.01, and 0.001.    

Within the observed scatter, the following features can be recognized:

\noindent - dSphs in the Local Group (\object{Sculptor}, \object{NGC~185}, and \object{NGC~205}) have a very low M$_{\rm HI}$/L$_B$ ratio ($\sim$ 0.001) compared with that of the dIrrs in groups (typically around unity) and of the field dIrrs ($\ge$ 1). 

\noindent - The dSph in the Sculptor group (\object{NGC~59}) has a ratio ($\sim$ 0.01) more similar to two field dSphs in the Millenium Galaxy strip (\object{MGC~0016030} [R12],  and \object{MGC~0038179} [R31]) with ratios of 0.06 and 0.15, respectively (Roberts et al.\ \cite{roberts}). 

\noindent - Also \object{Tucana} and \object{Cetus} (modulo the fact that we have only upper limits for the \HI) have a ratio of about 0.1, larger than other galaxies in the Local Group and more consistent with them being quasi-isolated.

\noindent - Although its \HI\ mass is just an upper limit, \object{APPLES~1} has an even higher ratio ($\sim$ 2), comparable to the field dSph \object{MGC~0060227} [R47]in the third Millenium Galaxy strip but, above all, it is also in agreement with the M$_{\rm HI}$/L$_B$ ratios of dIrrs in groups (open circles). Among the field dSphs and the dIrrs, the \HI\ mass of \object{APPLES~1} is the lowest.

As shown by Buyle et al. (\cite{buyle}; their Fig.~7), the dashed lines corresponding to M$_{\rm HI}$/L$_B$ of 2 and 1 are consistent with the theoretical predictions of a metal-poor (i.e. \object{LMC} and \object{SMC}) system undergoing self-regulating star formation (Pagel \& Tautvaisiene \cite{pagel}). This could be the case also for \object{APPLES~1} if its \HI\ content amounted to log(M$\rm _{\rm HI}$) $>$ 4.5 as determined from the observed scatter around these two lines. When fit using composite stellar populations following simple chemical enrichment models (PLF), the optical spectral energy distribution of \object{APPLES~1} suggests an extended ($\sim$ 10 Gyr) and weak ($\sim 10^{-3}$ M$_\odot$yr$^{-1}$) star formation history. Such star formation rate within $\sim 100$~pc -- which is the observed half-light radius of \object{APPLES~1} assuming a distance of 9.6 Mpc -- corresponds to  \HI\ gas masses $\sim 10^5$~M$_\odot$ for a standard Schmidt law (Kennicutt \cite{kennicutt}), i.e. an order of magnitude below the sensitivity limit of the observations presented here. The low efficiency of star formation is consistent with the low mass and size along with the isolated nature of this galaxy, which prevents any enhancement due to environmental effects as seen in local dwarf galaxies (Harris \& Zaritsky \cite{harris}). In this respect, then, \object{APPLES~1} would be confirmed as a dIrr, although gas-poor compared to other objects of the same class (typically with \HI\ masses M$\rm _{\rm HI}$ $>$ 10$^{7}$ \solmass; Fig.~2), where the primeval gas has been partly transformed into stars by a self-sustained and regulated star formation. However, given the relatively low gas mass, the optical morphology and star formation history, \object{APPLES~1} might, in our opinion, more credibly represent another example of the aforementioned transition-type dwarfs (e.\ g.\ \object{Phoenix} and \object{LGS~3}) with mixed characteristics (such as morphologies, stellar and gas masses, and angular momentum) of dIrrs and dSphs. In this case, the spheroidal morphology of \object{APPLES~1} could also be intrinsic and unaffected by the very low star formation rate or a recent interaction. Accordingly, the transitional phase of \object{APPLES~1} might represent a standard condition for `truly' isolated gas-poor dwarfs. Needless to say, this hypothesis, the origin of this class of dwarf galaxies, and its role in the evolutionary scenario between dwarf irregular and dwarf spheroidal/elliptical galaxies is still not yet fully investigated, mostly due to the few number of cases found.

Of course, together with the option that the gas has been and still is consumed by a low-level star formation activity, an interesting possibility is that part of the gas is ionized. The most recent episode of star formation (indicated by the presence of O and B stars), the potential presence of hot white dwarfs and UV-bright postasymptotic giant branch stars, and the SNIa heating (e.\ g.\ GGH and references therein) could all contribute to ionize a fraction of the whole \HI\ content of \object{APPLES~1}. Therefore, the mass of the neutral gas left may fall below the sensitivity of our observations. The surface brightness of this ionized gas could be so low to have escaped detection in the optical spectra acquired by PLF.

\section{Summary}

We performed \HI\ sensitive observations of the field dwarf galaxy \object{APPLES~1} with the Parkes radiotelescope, the main result of which is a stringent upper limit on the neutral hydrogen gas mass of the galaxy, i.e. $M_{\rm HI}$ $<$ $10^{6}$ \solmass, and, as a consequence, a value for the M$_{\rm HI}$/L$_{\rm B}$ ratio smaller than 2.4. We have compared these values with with those of other known dwarf galaxies, either in the field or in groups deducing that \object{APPLES~1} is likely a further example of dIrr/dSph transition type since it shows mixed charachteristics of the two classes: somewhat spherical morphology and relatively low gas mass like in dSphs but also recent star formation typical of dIrrs. We conclude that an extended and low-efficiency star formation activity can explain the neutral gas content found in \object{APPLES~1}. The star-formation history of \object{APPLES~1} might itself able to ionize part of the neutral gas, so that the total effect is to reduce the \HI\ content below the detection threshold of the Parkes telescope. The study of dwarf field galaxies like \object{APPLES~1} is important since, being not influenced by external mechanisms, they are unique objects to test theories of star formation and the evolution of galaxies, with particular attention to the evolutionary path from dwarf irregular to dwarf spheroidal/elliptical galaxies. Of course, to really exploit the information derivable from these types of galaxies, surveys to search for (faint) field dwarf galaxies are needed.

\begin{acknowledgements}
We wish to thank Mario Mateo, the referee, for his comments on the manuscript and Oleg Gnedin for useful suggestions. This research has made use of NASA's Astrophysics Data System. This research has made use of the NASA/IPAC Extragalactic Database (NED) which is operated by the Jet Propulsion Laboratory, California Institute of Technology, under contract with the National Aeronautics and Space Administration.
\end{acknowledgements}


\begin{thebibliography}{}
 
\bibitem[2005]{bouchard}
Bouchard, A., JerJen, H., Da Costa, G.S., Ott, J., 2005, AJ, accepted
\bibitem[2005]{buyle}
Buyle, P., De Rijcke, S., Michielsen, D., Baes, M., Dejonghe, H., 2005, MNRAS, accepted, astro-ph/0504270
\bibitem[1991]{carignan}
Carignan, C., Demers, S., Cot\'{e}, S., 1991, ApJ 381, 13
\bibitem[2003]{grebel}
Grebel, E.K., Gallagher, J.S., III, Harbeck, D., 2003, AJ 125, 1926; GGH
\bibitem[2004]{harris}
Harris, J. \& Zaritsky, D.\ 2004, AJ 127, 1531
\bibitem[2000]{huchtmeier}
Huchtmeier, Karachentsev, I.D., Karachentseva, V.E., 2000, A\&AS 147, 187
\bibitem[2005]{karachentsev1}
Karachentsev, I.D., 2005, AJ, 129, 178
\bibitem[2001]{karachentsev2}
Karachentsev, I.D., Karachentseva, V.E., Huchtmeier, W.K., 2001, A\&A 366, 428
\bibitem[1998]{kennicutt}
Kennicutt, R.~C.\ 1998, ApJ 498, 541
\bibitem[1999]{klypin}
Klypin, A., Kravtzov, A.V., Valenzuela, O., et al., 1999, ApJ 522, 82
\bibitem[1993]{lo}
Lo, K.Y., Sargent, W.L.W., Young, K., 1993, AJ 106, 507
\bibitem[1998]{mateo}
Mateo, M.L., 1998, ARA\&A 36, 435
\bibitem[2005]{mayer}
Mayer, L., Mastropietro, C., Wadsley, J., Stadel, J., Moore, B., 2005, MNRAS, astro-ph/0504277
\bibitem[1996]{oosterloo}
Oosterloo, T., Da Costa, G.S., Staveley-Smith, L., 1996, AJ 112, 1969
\bibitem[1998]{pagel}
Pagel, B.E.J., Tautvaisiene, G., 1998, MNRAS 299, 535
\bibitem[2005]{pasquali} 
Pasquali, A, Larsen, S., Ferreras, I., et al., 2005, AJ 129, 148; PLF
\bibitem[2003]{paturel}
Paturel, G., Theureau, G., Bottinelli, L., et al., 2003, A\&A, 412, 57
\bibitem[2004]{roberts}
Roberts, S., Davies, J., Sabatini, S. et al., 2004, MNRAS 352, 478
\bibitem[1991]{sandage}
Sandage, A. \& Hoffman, G.L.\ 1991, ApJ 379, 45
\bibitem[2002]{stil}
Stil, J.M., Israel, F.P., 2002, A\&A 389, 42
\bibitem[1994]{vandenbergh}
van den Bergh, S., 1994, ApJ 428, 617
\bibitem[2001]{vanzee}
van Zee, L., 2001, AJ 121, 2003

\end{thebibliography}
\end{document}